\documentclass[aps,twocolumn,superscriptaddress]{revtex4}
\usepackage{graphicx}
\usepackage{amssymb}
\usepackage{amsmath}
\usepackage{epsfig}
\usepackage{color}
\usepackage{verbatim}
\usepackage{float}
\usepackage{mathrsfs}
\usepackage{esint}
\usepackage{tabularx}
\usepackage{multirow}
\usepackage{braket}
\usepackage{color,soul}
\usepackage{mathtools}

\usepackage{array}
\newcolumntype{C}[1]{>{\centering\let\newline\\\arraybackslash\hspace{0pt}}m{#1}}

\begin{document}

\title{Excitations and impurity dynamics in a fermionic Mott insulator with nearest-neighbor interactions}
\author{A.-M. Visuri}
\affiliation{COMP Centre of Excellence, Department of Applied Physics, Aalto University, FI-00076 Aalto, Finland}
\affiliation{Department of Quantum Matter Physics, University of Geneva, 24 quai Ernest-Ansermet, 1211 Geneva, Switzerland}

\author{T. Giamarchi}
\affiliation{Department of Quantum Matter Physics, University of Geneva, 24 quai Ernest-Ansermet, 1211 Geneva, Switzerland}

\author{P. T\"{o}rm\"{a}}
\email{paivi.torma@aalto.fi}
\affiliation{COMP Centre of Excellence, Department of Applied Physics, Aalto University, FI-00076 Aalto, Finland}
\affiliation{Institute for Quantum Electronics, ETH Zurich, 8093 Zurich, Switzerland}

\begin{abstract}

We study analytically and with the numerical time-evolving block decimation method the dynamics of an impurity in a bath of spinless fermions with nearest-neighbor interactions in a one-dimensional lattice. The bath is in a Mott insulator state with alternating sites occupied and the impurity interacts with the bath by repulsive on-site interactions. We find that when the magnitudes of the on-site and nearest-neighbor interactions are close to each other, the system shows excitations of two qualitatively distinct types. For the first type, a domain wall and an anti-domain wall of density propagate into opposite directions, while the impurity stays at the initial position. For the second one, the impurity is bound to the anti-domain wall while the domain wall propagates, an excitation where the impurity and bath are closely coupled.

\end{abstract}

\maketitle

\section{Introduction}

A single particle, or a macroscopic quantum degree of freedom, coupled to a bath is a paradigmatic problem of many-body physics. It is naturally described, at least in three dimensions, as a composite object of the particle dressed by bath excitations, which is a quasiparticle. Dressing renormalizes some of the particle's properties such as the mass. This concept has been particularly fruitful for the polaron problem, where the bath consists of phonons \cite{Feynman,grusdt_polaron_review}, and for the Fermi-liquid theory where the collective action of all the indiscernible particles in an interacting Fermi gas renormalizes the parameters of a single-particle excitation \cite{nozieres_book}. In certain cases, such as the Caldeira-Leggett problem, the renormalization of the parameters can be strong enough to significantly change the behavior of the particle \cite{leggett_two_state}.

Similar effects occur in restricted geometries such as one-dimensional quantum systems where the effects of interactions are considerably reinforced. In particular, it was shown that an impurity can behave quite differently from a quasiparticle and the interaction with the bath can lead to subdiffusion \cite{zvonarev_ferro_cold}. This new phenomenon has triggered an intense theoretical and experimental activity, in particular with very controlled experimental realizations by cold atomic gases. Optical lattices in experiments with ultracold gases are devoid of phonons, and in order to study similar phenomena as in condensed matter physics, phonons can be incorporated via a bath of a different type of particles \cite{Widera, Bruun}. Mobile impurities in fermionic and bosonic baths have recently been studied to a large extent theoretically \cite{Jaksch, Kantian, Sirker} and experimentally \cite{Kohl, Catani, Fukuhara}. In one dimension in particular, interesting time-dependent phenomena have been predicted, such as a crossover from a bound molecule to a polaron \cite{Massel}, the damping of Bloch oscillations \cite{Jaksch, Kamenev, Grusdt}, the non-relaxation of a supersonic impurity \cite{Demler}, and an intriguing behavior of pair correlations with slow and fast driven barriers \cite{Visuri}.

In the aforementioned studies, the bath is assumed to be homogeneous, and the only deformations of the bath are caused by
its fluctuations and the interaction with the impurity. This restriction is quite natural for systems with a contact interaction
such as cold atoms. In this article, we address the question of how an impurity behaves if the bath instead possesses an internal structure, such as a periodic arrangement of the bath particles. An impurity would see an external periodic potential but would also be able to create excitations, a situation not dissimilar to the presence of phonons and the electron-phonon coupling in a solid.  Such baths can for instance be realized by long-range interactions.

We investigate the dynamics of an impurity interacting with a bath of fermions with nearest-neighbor interactions. Using the numerical time-evolving block decimation (TEBD) method \cite{Vidal, Daley} and analytic arguments, we show that the dynamics of the impurity drastically changes when the impurity can create excitations in the bath, leading to a bound state of the impurity and distortions in the bath. Nearest-neighbor as well as longer-range interactions are currently becoming feasible in experiments with ultracold bosonic \cite{DipolarBosonReview, Nagerl, Kotochigova, EBHM_Ferlaino2015} and fermionic \cite{Ye, Lev, Zwierlein, Jin} dipolar molecules, dipolar atoms and Rydberg atoms \cite{Raithel, Pfau} both in traps and lattices. These advances make setups available for testing the predictions presented here. The motion of an impurity in an optical lattice can be recorded by single-site-resolved imaging \cite{Fukuhara}. Previous theoretical studies have considered spinless fermions with nearest-neighbor interactions in the context of interaction quenches \cite{Muramatsu}, and the excited states created inside the Mott gap of a one-dimensional solid by applying a laser pulse \cite{Cavalleri, Tohyama, Prelovsek}, described by the extended Hubbard model with nearest-neighbor interactions among two spin species. Such excitations are delocalized and have a well-defined energy unlike the initially localized domain wall excitations studied here.

The model and the numerical method are introduced in Section \ref{section:model}. In Section \ref{section:doublons}, we show how the number of doublons evolves in time for different regimes of interactions and explain the case $V \gg U$ by an analytic model for the $V \rightarrow \infty$ limit. The short-time dynamics are modeled by a three-site Hamiltonian in Section \ref{section:three-site}. Section \ref{section:excitations} discusses the excitation dynamics in detail. We illustrate different possible excitation processes which explain the features observed in density distributions. In order to find a more precise picture of how the bath evolves in time, we also study the correlation of density differences as a function of the distance from the center of the lattice. These results are presented in Section \ref{section:density_correlation}. The possible experimental realization of the model with ultracold dipolar gases is discussed in Section \ref{section:experiments}. Finally, a brief summary is presented in Section \ref{section:conclusions}.

\section{The model and the numerical method}
\label{section:model}

\begin{figure}[h!]
\begin{center}
\includegraphics[width=\linewidth]{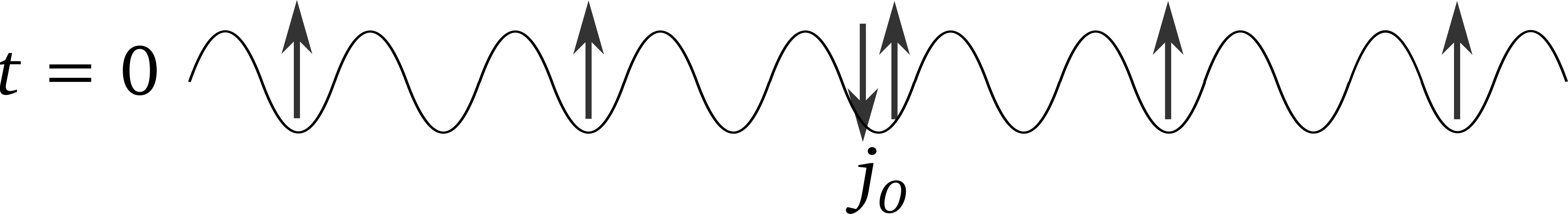}
	\caption{A schematic figure of the initial configuration where the total number of doubly occupied sites is $\langle N_{\uparrow\downarrow}(t=0) \rangle = 1$.}
\label{fig:initial_configuration}
\end{center}	
\end{figure}

The impurity and the bath are described by the Hamiltonian 
\begin{align*}
H = &- J \sum_{j\sigma} (c_{j\sigma}^{\dagger} c_{j+1\sigma} + \text{h.c.}) \\ 
&+ U \sum_j n_{j\uparrow} n_{j\downarrow} + V \sum_j n_{j\uparrow} n_{j+1\uparrow}.
\end{align*}
Here, $c_{j\uparrow}$ annihilates a bath fermion and $c_{j\downarrow}$ the impurity, and $n_{j\sigma} = c_{j\sigma}^{\dagger}c_{j\sigma}$ is the number operator. The tunneling energy is denoted by $J$. Opposite spins interact on the same site with energy $U > 0$, and the bath fermions among nearest neighbors  with $V > 0$. In the initial state, $U = 0$ and the impurity is localized at the center of the lattice (site $j_0$), as shown in Fig. \ref{fig:initial_configuration}. At half-filling and $V > 2J$, the ground state of the bath is a Mott insulator \cite{Giamarchi}. In order to find a non-degenerate ground state, we fix the number of lattice sites to $L = 2p + 1$, where $p \in \mathbb{N}$, and the number of bath fermions to $N_{\uparrow} = p + 1$. For even $p$, $j_0$ is occupied by a bath fermion, and for odd $p$, $j_0$ is empty. In our TEBD calculations, $L$ varies from 79 to 81, the Schmidt number in the truncation of the state is fixed to 100, and a time step $0.02\frac{1}{J}$ is used in the real time evolution.

\section{The time evolution of the number of doublons}
\label{section:doublons}

\begin{figure}[h!]
\begin{center}
  \hspace{-0.5cm}
  \begin{minipage}[l]{0.08\linewidth}
    a)
  \end{minipage}
  \begin{minipage}[c]{0.9\linewidth}
    \includegraphics[width=\textwidth]{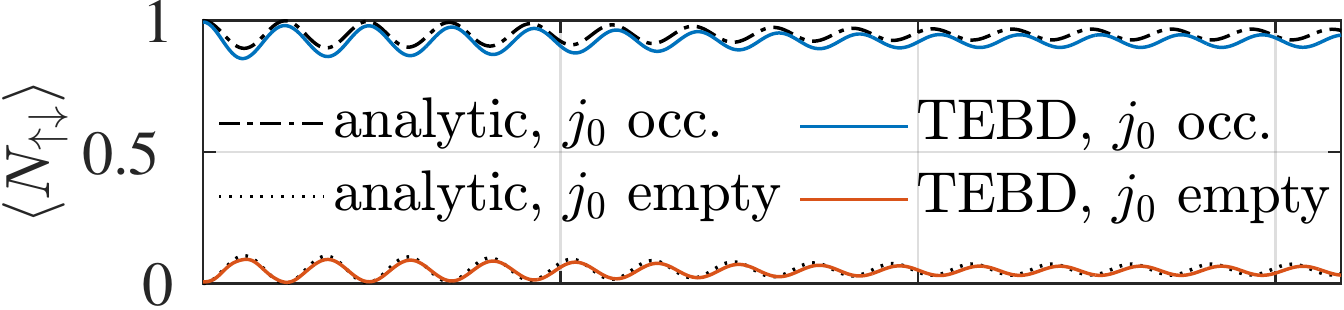}
  \end{minipage}
  
  \hspace{-0.5cm}
  \begin{minipage}[l]{0.08\linewidth}
    b)
  \end{minipage}
  \begin{minipage}[c]{0.9\linewidth}
    \includegraphics[width=\textwidth]{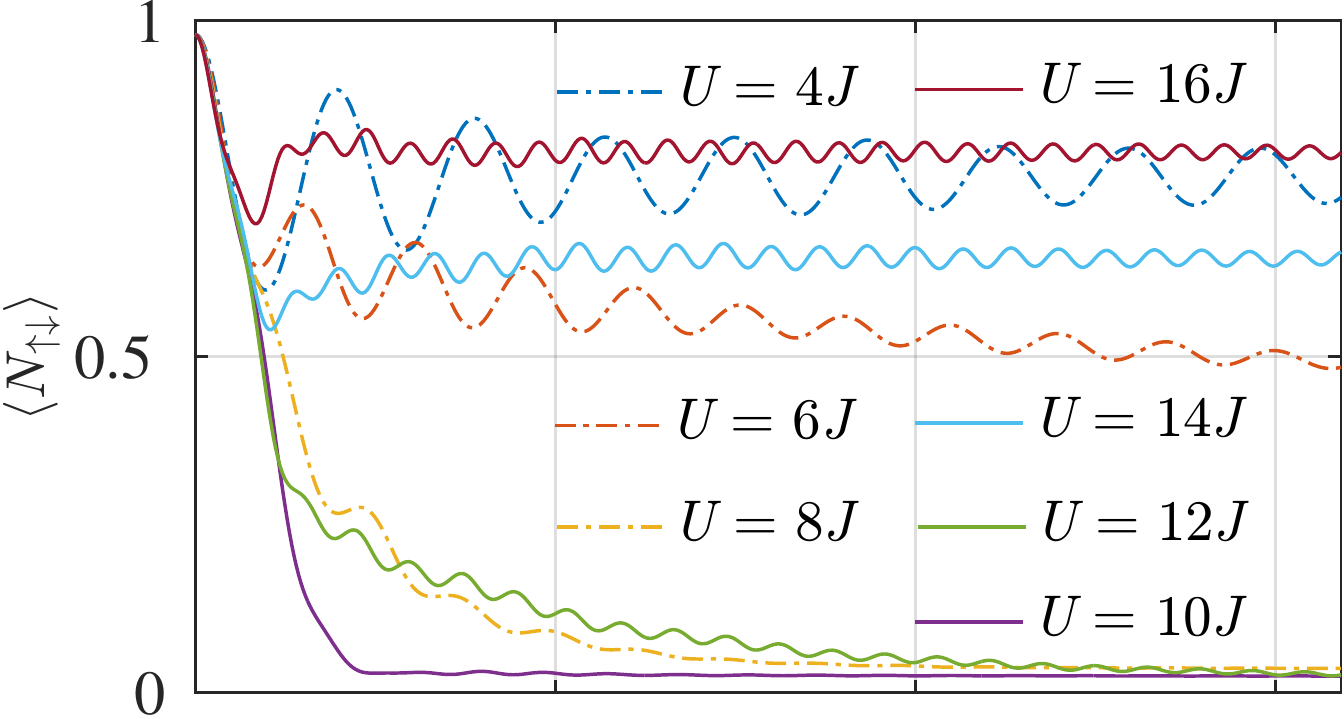}
  \end{minipage}
  
  \hspace{-0.5cm}
  \begin{minipage}[l]{0.08\linewidth}
    c)
  \end{minipage}
  \begin{minipage}[c]{0.9\linewidth}
    \includegraphics[width=\textwidth]{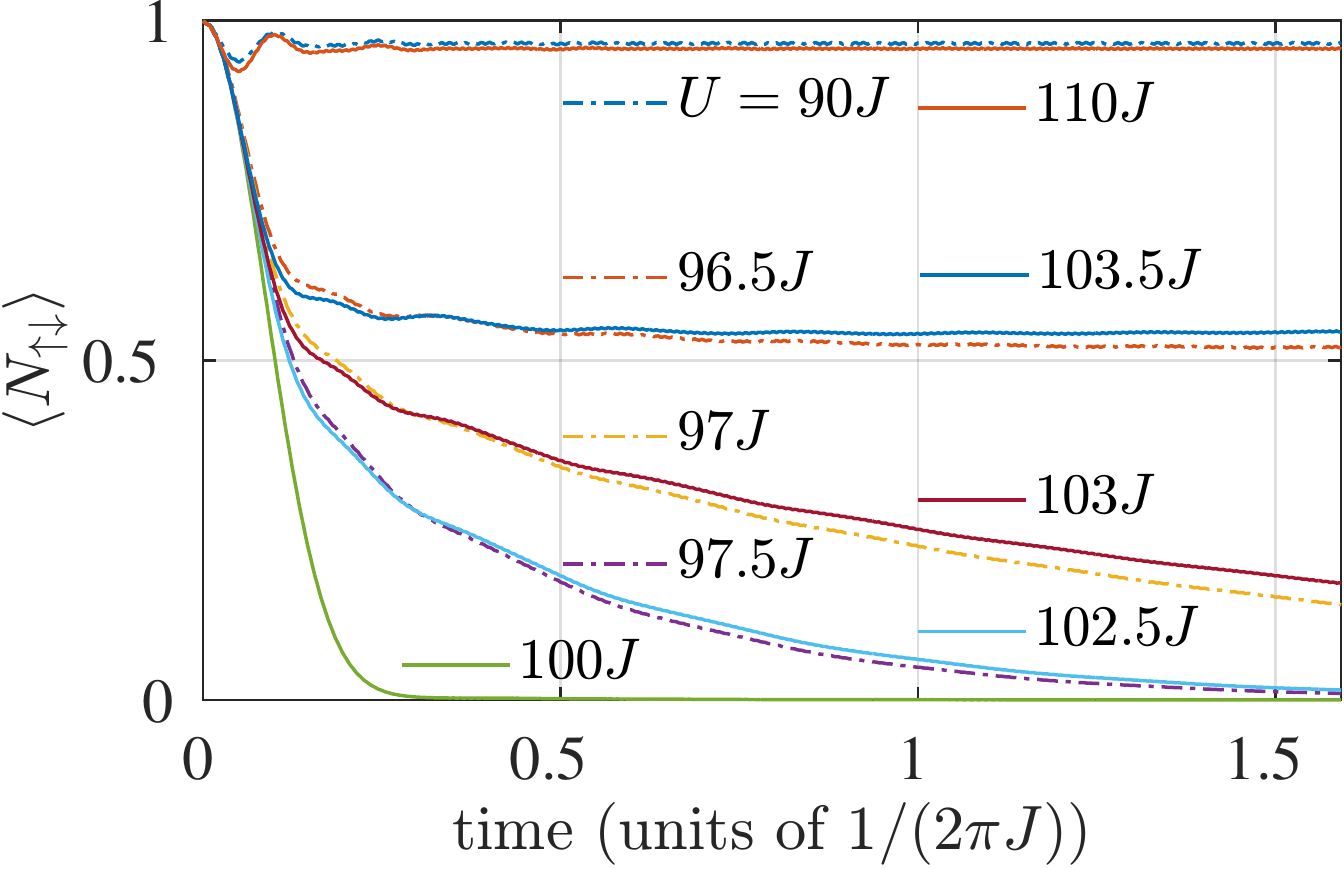}
  \end{minipage}
	\caption{a) The number of doubly occupied sites $\langle N_{\uparrow\downarrow}(t) \rangle$ calculated analytically in the $V \rightarrow \infty$ limit for $U = 8J$, and by TEBD for $U = 8J$ and $V = 20J$. Here, $V \gg U$ and the number of doublons is nearly conserved. b) TEBD results for $V = 10 J$ and varying $U$. For $U$ sufficiently close to $V$, $\langle N_{\uparrow\downarrow}(t) \rangle$ decays as a function of time. c) TEBD results for $V = 100J$ and varying $U$. }
\label{fig:doublons_V100}
\end{center}	
\end{figure}

In the beginning of the time evolution, the on-site interaction is switched to $U > 0$ and the impurity is released. Due to energy conservation, the dynamics will be different for $U$ close to $V$ -- at resonance -- and for $U$ and $V$ far detuned. Off resonance, with $U \gg J$, the total number of doubly occupied sites $\langle N_{\uparrow\downarrow}(t) \rangle = \bra{\psi(t)} \sum_j n_{j\uparrow} n_{j\downarrow} \ket{\psi(t)}$ oscillates with a frequency close to $U$ while the average value stays constant, as seen in Fig. \ref{fig:doublons_V100} a). The behavior agrees well with the analytic solution for a free particle in a superlattice with a potential difference $U$ between alternating sites, which is equivalent to the impurity problem when $U \gg J$ and $V \rightarrow \infty$ (see Appendix \ref{app:superlattice}). Figure \ref{fig:doublons_V100} a) shows that the analytic model describes the impurity problem well for an impurity created at either an occupied or an empty site. The oscillation frequency $U$ is also seen for the parameters of Fig. \ref{fig:doublons_V100} b) when $|V - U|$ is sufficiently large. In Fig. \ref{fig:doublons_V100} c), the frequency is very high and the amplitude small and therefore the oscillation is almost invisible. It can be seen at a shorter time scale in Section \ref{section:three-site}. 

For an impurity created at an occupied site with $U$ and $V$ far detuned, the energy $U$ cannot be deposited into the bath. We find that the impurity propagates on occupied sites in a second-order process with velocity $\frac{4J^2}{U}$, which is the superexchange coupling obtained from a mapping to the Heisenberg Hamiltonian at the $U \gg J$ limit \cite{Georges}. Similarly, if the impurity starts at an empty site, it will not have enough energy to move to a site occupied by a bath fermion and will propagate on the empty sites. In contrast, an impurity at an occupied site with $U \approx V$ can deposit the energy $U$ into the bath for example in a process where the bath fermion at $j_0$ moves by one site. This is seen as a decay of $\langle N_{\uparrow\downarrow}(t) \rangle$ in Figs. \ref{fig:doublons_V100} b) and c).

It is of interest to ask what kind of excitations are created in this process. To unambiguously investigate the basic types of excitations in the bath, we focus on the case of Fig. \ref{fig:doublons_V100} c) with very large $U$ and $V$ which suppresses pair tunneling processes. In Fig. \ref{fig:doublons_V100} c), $V$ is fixed to $100J$ and $U$ varied around this value. Intriguingly, we find that the curves for which  $|V - U| \geq 3.5J$ saturate to a nonzero constant whereas the ones for which $|V - U| \leq 2.5J$ decay to a value close to zero. For $|V - U| = 3J$, there is a decay within the simulation time but it is unclear whether $\langle N_{\uparrow\downarrow} \rangle$ approaches zero at longer times. These results are discussed in detail in Section \ref{section:excitations}. Essentially the same behavior is obtained for $V = 10 J$ in Fig. \ref{fig:doublons_V100} b), which is closer to experimentally realizable values, as discussed in Section \ref{section:experiments}.

\section{Three-site model for the short-time dynamics}
\label{section:three-site}

\begin{figure}[h!]
\begin{center}
	\includegraphics[width=0.65\linewidth]{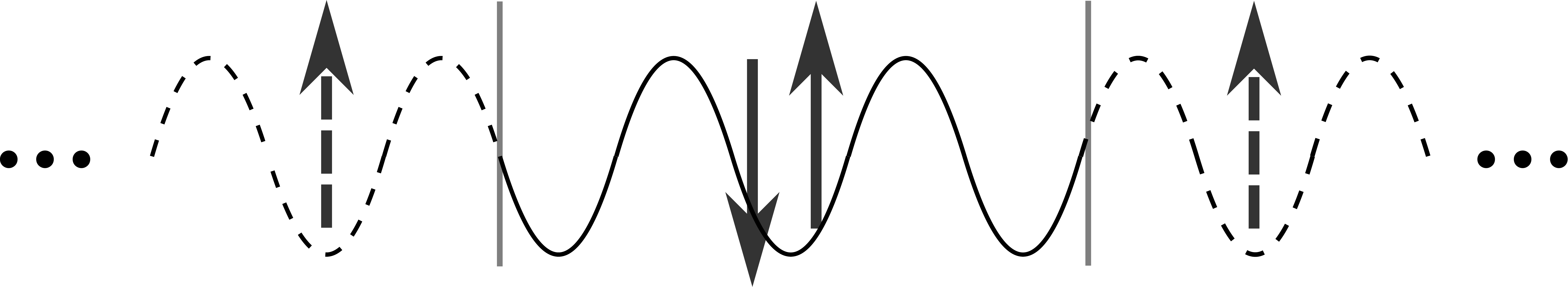}
	\caption{The three-site model comprises of the three central sites between the vertical lines.}
\label{fig:three_sites}
\end{center}	
\end{figure}

For the large $V = 100 J$ of Fig. \ref{fig:doublons_V100} c), the number of doublons for $U = 90J$ and $U = 110J$ has an initial oscillation with frequency close to $|V - U|$, in addition to the high frequency $U$ explained in Section \ref{section:doublons}. A similar initial behavior is given by a three-site model, illustrated in Fig. \ref{fig:three_sites}, where $N_{\uparrow} = N_{\downarrow} = 1$. The two particles have an on-site interaction $U$, and the spin-up particle has a potential $V$ at sites 1 and 3 mimicking fixed spin-up fermions at the adjacent sites. In the basis
\begin{align*}
\ket{1} &= \ket{\uparrow\downarrow \: \: \emptyset \: \: \: \emptyset}, 
&\ket{4} &= \ket{\uparrow \: \: \: \downarrow \: \: \: \emptyset},
&\ket{7} &= \ket{\emptyset \: \: \: \downarrow \: \: \: \uparrow}, \\
\ket{2} &= \ket{\emptyset \: \: \uparrow\downarrow \: \: \emptyset}, 
&\ket{5} &= \ket{\downarrow \: \: \: \uparrow \: \: \: \emptyset},
&\ket{8} &= \ket{\uparrow \: \: \: \emptyset \: \: \: \downarrow}, \\
\ket{3} &= \ket{\emptyset \: \: \: \emptyset \: \: \uparrow\downarrow}, 
&\ket{6} &= \ket{\emptyset \: \: \: \uparrow \: \: \: \downarrow}, 
&\ket{9} &= \ket{\downarrow \: \: \: \emptyset \: \: \: \uparrow}, \\
\end{align*}
the Hamiltonian is written as
\begin{align*}
H = 
\begin{pmatrix} 
U + V	&0		&0		&-t		&-t		&0		&0		&0		&0	\\ 
0		&U		&0		&-t		&-t		&-t		&-t		&0		&0	\\
0		&0		&U + V	&0		&0		&-t		&-t		&0		&0	\\
-t		&-t		&0		&V		&0		&0		&0		&-t		&0	\\
-t		&-t		&0		&0		&0		&0		&0		&0		&-t	\\
0		&-t		&-t		&0		&0		&0		&0		&-t		&0	\\
0		&-t		&-t		&0		&0		&0		&V		&0		&-t	\\
0		&0		&0		&-t		&0		&-t		&0		&V		&0	\\
0		&0		&0		&0		&-t		&0		&-t		&0		&V
\end{pmatrix}.
\end{align*}
The initial state is a superposition of states $\ket{2}$, $\ket{4}$, and $\ket{7}$.

\begin{figure}[h!]
\begin{center}
	\includegraphics[width=\linewidth]{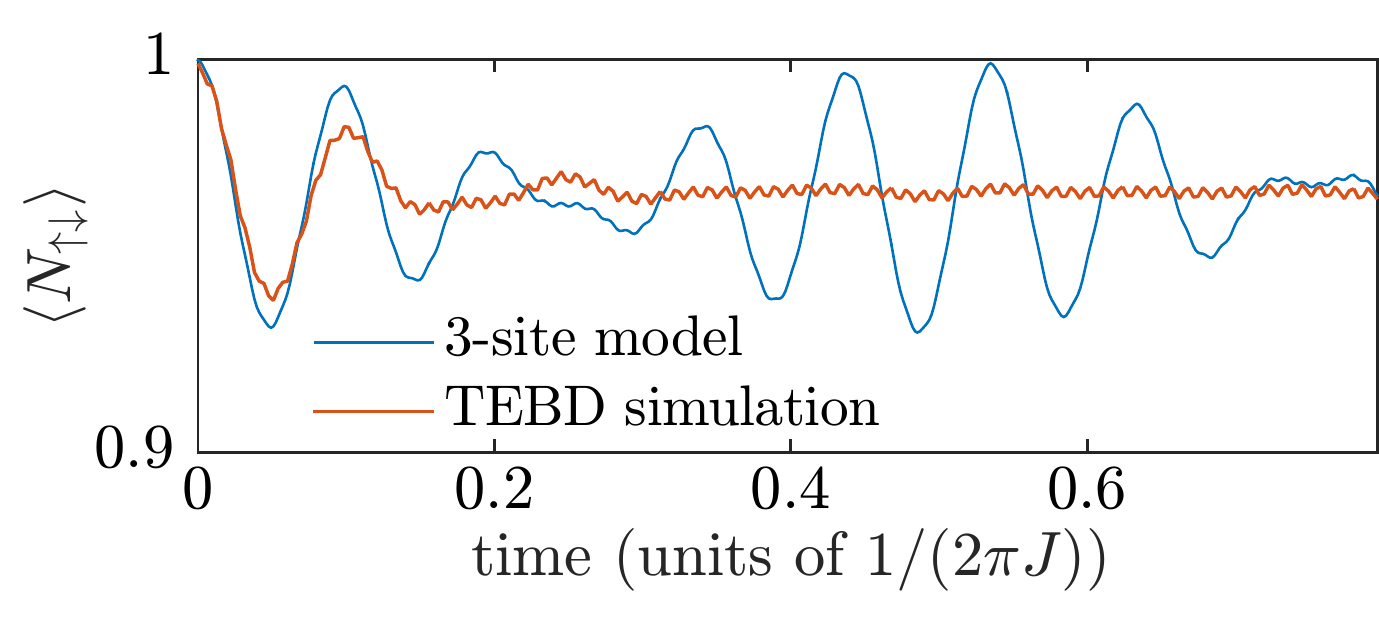}
	\caption{The total number of doublons $\langle N_{\uparrow\downarrow}(t) \rangle$ for $U = 90 J$ and $V = 100 J$ has a similar initial behavior in the three-site model as in the many-body TEBD simulations.}
\label{fig:doublons_HT_V100}
\end{center}	
\end{figure}

In the three-site model, the initial behaviour of $\langle N_{\uparrow\downarrow}(t) \rangle$ during the first oscillation period is close to the many-body result, as shown in Fig. \ref{fig:doublons_HT_V100}. For three sites, there is a revival of the oscillation after damping, which is not seen in the many-body case. This indicates that the excitations propagating away from the three central sites play a role in the dynamics after the initial stage. One can therefore conclude that the bath fermions at sites beyond the neighboring ones are responsible for the permanent damping. The average value to which $\langle N_{\uparrow\downarrow} \rangle$ saturates agrees with the three-site model.

\section{Excitations in the bath}
\label{section:excitations}

\subsection{Particle densities}

\begin{figure}[h!]
\begin{center}
	\includegraphics[height = 0.35\linewidth]{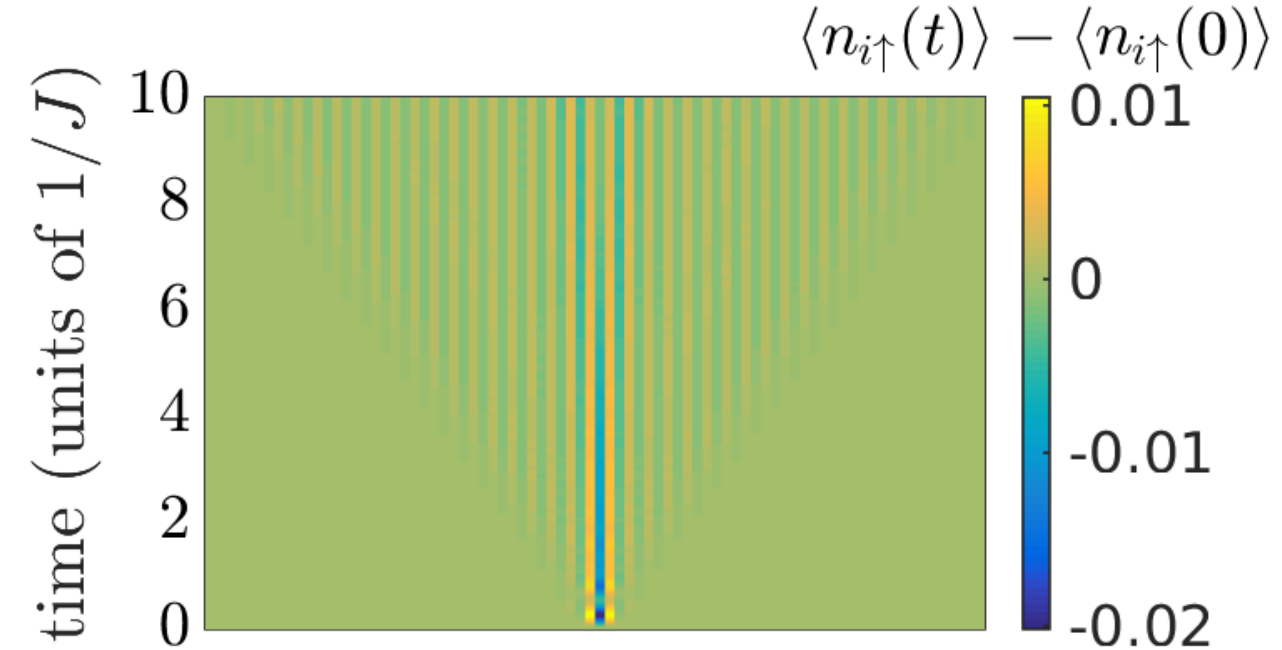}
	\includegraphics[height = 0.35\linewidth]{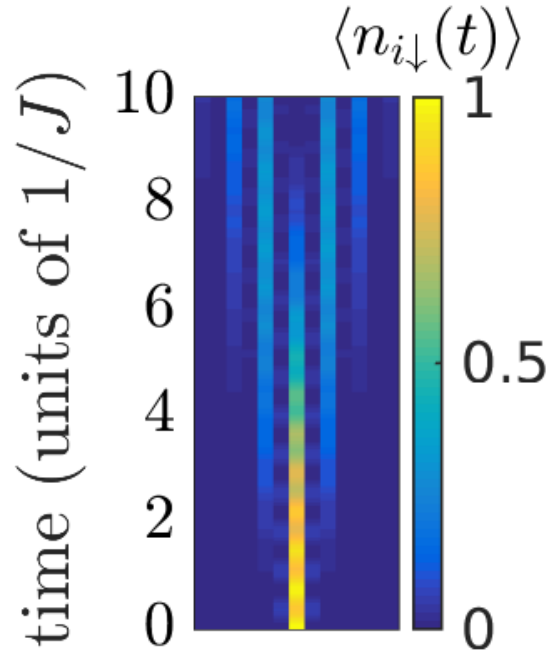}
	
	\includegraphics[height = 0.38\linewidth]{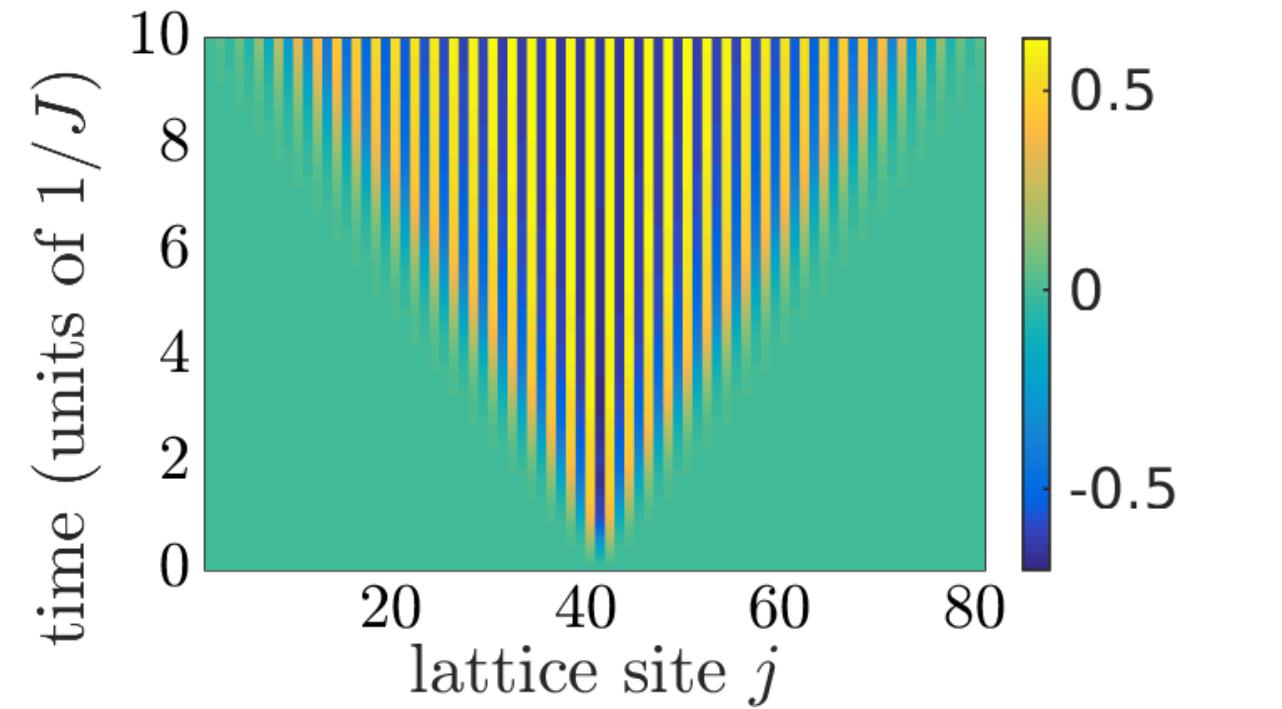}
	\includegraphics[height = 0.38\linewidth]{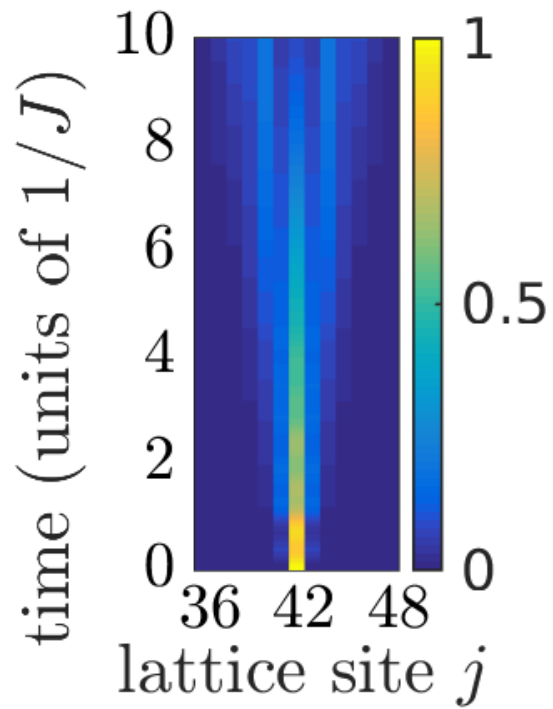}	
	\caption{Left column: The difference in the density of spin-up (bath) fermions with respect to the initial state as a function of position and time, $\bra{\psi(t)} n_{j\uparrow} \ket{\psi(t)} - \bra{\psi(0)} n_{j\uparrow} \ket{\psi(0)}$ for $U = 8 J$ and $V = 20 J$ (upper row) and $U = 8 J$ and $V = 10 J$ (lower row). Right column: The density of the spin-down impurity $\bra{\psi(t)} n_{j\downarrow} \ket{\psi(t)}$ for the same values of $U$ and $V$. For $U$ close to $V$, the changes in the bath density distribution (lower left panel) are an order of magnitude larger than for $U$ and $V$ far apart (upper left panel), indicating that there is a resonant interval of $U$ around $V$ for creating excitations. }
\label{fig:densities_supplemental}
\end{center}
\end{figure}

\begin{figure}[h!]
\begin{center}
	\includegraphics[height=0.35\linewidth]{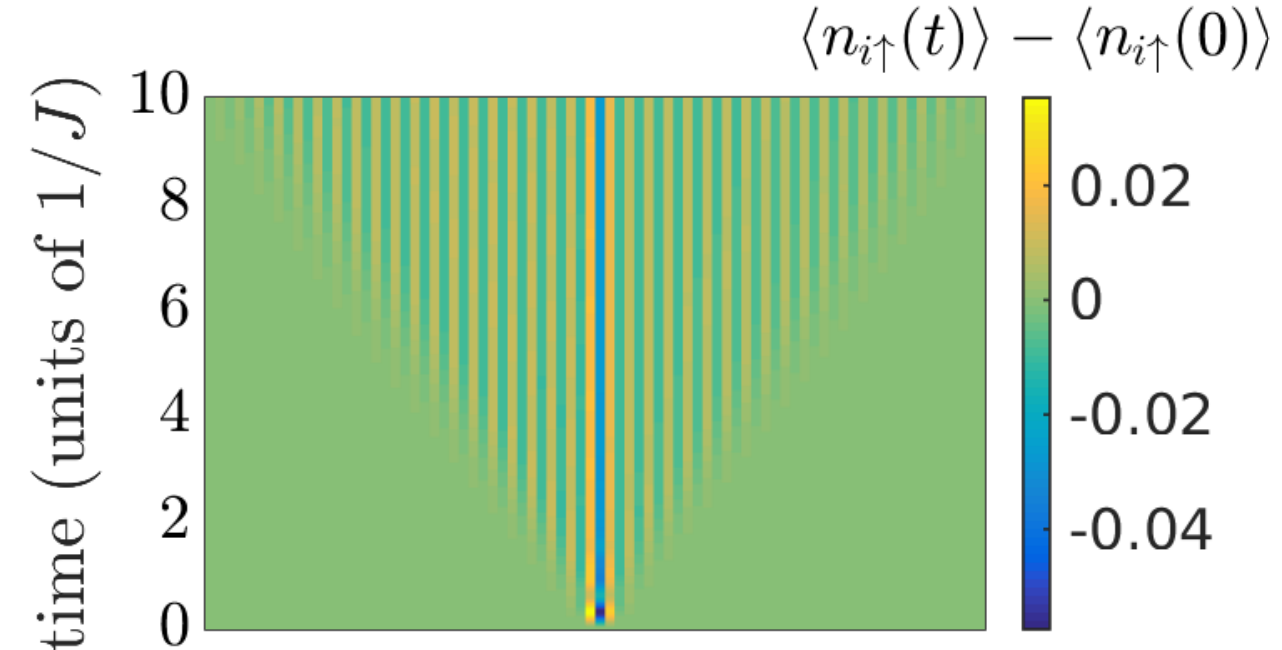}
	\includegraphics[height=0.35\linewidth]{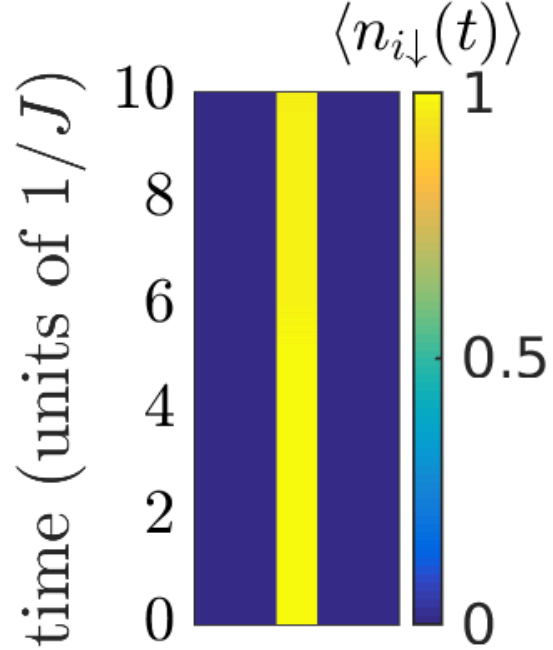}
	
	\includegraphics[height=0.38\linewidth]{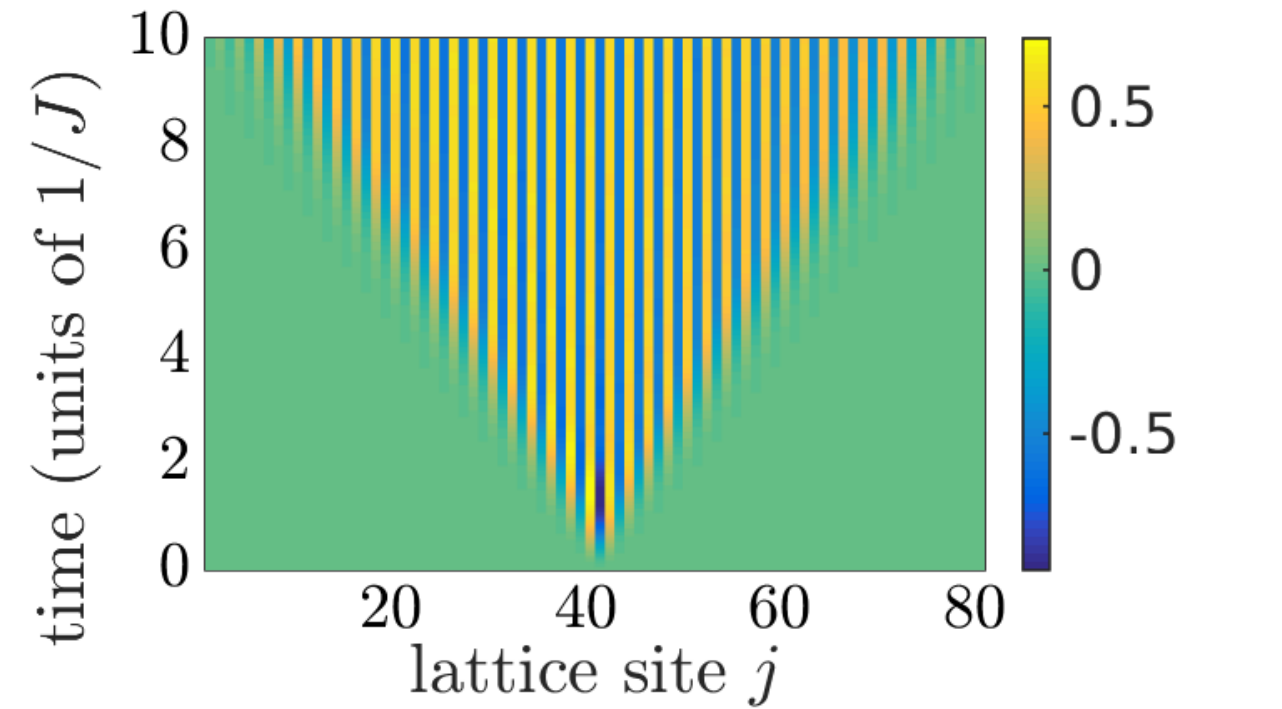}
	\includegraphics[height=0.38\linewidth]{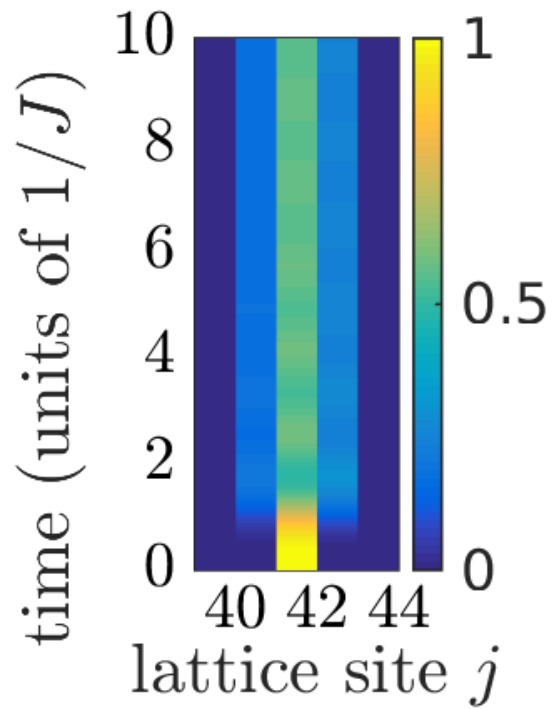}
	\caption{The same quantities as in Fig. \ref{fig:densities_supplemental} for $V = 100 J$ and $U = 90J$ (upper), $U = 100J$ (lower). The parameters of the upper row are in the off-resonant region and those of the lower row in the resonant region.}
\label{fig:densities}
\end{center}	
\end{figure}

The decay of $\langle N_{\uparrow\downarrow} \rangle$ for $U$ close to $V$ is connected to the creation of excitations in the bath. The excitations can be seen as a density difference $\langle n_{j\uparrow}(t) \rangle - \langle n_{j\uparrow}(0) \rangle$ propagating from the center of the lattice in Figs. \ref{fig:densities_supplemental} and \ref{fig:densities}. In the off-resonant cases, the density differences are an order of magnitude smaller than in the resonant cases, and the impurity propagates more diffusively at resonance. Off resonance, the propagation is limited to occupied sites, which is a second-order process with velocity $\frac{4 J^2}{U}$ \cite{Georges}. In Fig. \ref{fig:densities_supplemental}, the value $U = 8 J$ allows a propagation velocity of the impurity which can be observed within the simulation time, whereas for $U = 90 J$ in Fig. \ref{fig:densities} the velocity is too small to be observed.
\subsection{Model for the resonance region}
\label{section:energy_conservation}

The density differences $\langle n_{j\uparrow}(t) \rangle - \langle n_{j\uparrow}(0) \rangle$ are largest for $U = V$. To explain this behavior, one can look for limits of $|V - U|$ within which excitations can be created. For $V \rightarrow \infty$, the mapping to a free particle in a superlattice gives two energy bands for the impurity, $\pm E_k = \pm \sqrt{4J^2\cos^2(k) + (\frac{U}{2})^2}$, where $k$ is the quasi-momentum (see Appendix \ref{app:superlattice}). An impurity created at an occupied site is initially in the higher band. A simple way to derive the width of the resonance in $U$ is to assume that the impurity transitions to the lower band by moving to an empty site, and that the dispersion relation of the impurity is unchanged in the transition. The energy released in such a transition would be absorbed by excitations created in the bath. In this scenario, the excitations in the bath are created far away and do not interact with the impurity. In such an excitation, a bath fermion moves by one site, which corresponds to creating a domain wall (DW) with two neighboring sites occupied and an anti-domain wall (ADW) with two neighboring sites empty. The bath can be mapped to an XXZ Hamiltonian, which has a two-domain-wall excitation continuum $\omega \in [V - 4J, V + 4J]$ \cite{Shiba, Kolezhuk}. Assuming the same excitation spectrum as for two domain walls \cite{Shiba, Kolezhuk}, the process of creating a DW and an ADW has minimally the energy cost $V - 4J$ and maximally $V + 4J$. By energy conservation, the lower limit of $U$ to create excitations is $U_{\min} = \sqrt{(V - 4J)^2 - 16J^2}$ and the upper limit is $U_{\max} = V + 4J$ (see Appendix \ref{app:resonance_region}). For $V = 100J$, $U_{\min} \approx 95.9J$ and $U_{\max} = 104J$.

One can see in Fig. \ref{fig:doublons_V100} c) that for $U = V \pm 2.5J$, $\langle N_{\uparrow\downarrow} \rangle$ decays to a value close to zero and for $U = V \pm 3J$, there is a slower decay. On the other hand, for $U = V \pm 3.5J$ which is within $[U_{\min}, U_{\max}]$, $\langle N_{\uparrow\downarrow} \rangle$ saturates to approximately 0.5. As seen in Fig. \ref{fig:doublons_V100} b), a saturation to a value less than one also occurs in the analytic superlattice model where there can be no excitations. This suggests that creating a DW-ADW excitation far from the impurity does not have a high probability since the clear decay behavior does not persist to $U = V \pm 4J$. Thus the bounds $U_{\min}$ and $U_{\max}$ derived above do not describe the numerical results accurately. It is notable and curious that this straightforward description does not agree with the simulations.

\subsection{Excitation processes and the observed resonance region}

\begin{figure*}[t!]
\begin{center}
	\includegraphics[width=0.8\textwidth]{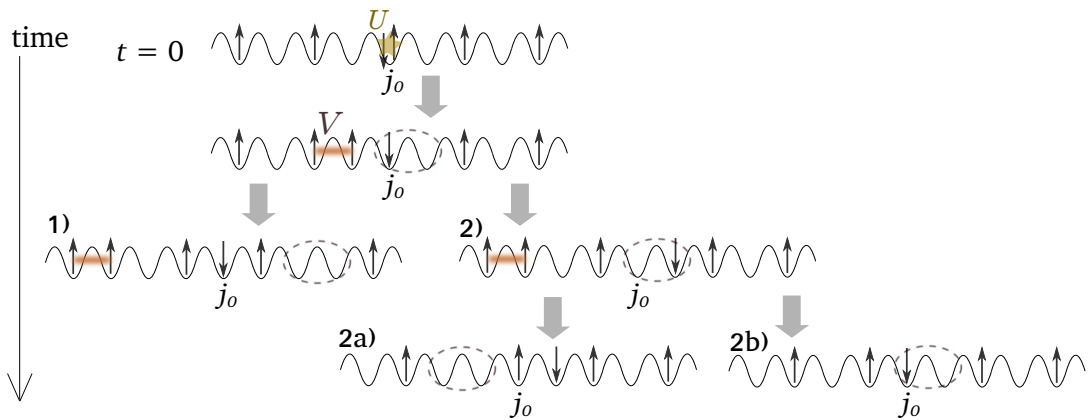}
	\caption{A schematic figure of the time evolution where a DW-ADW excitation is created by moving a bath particle to the neighboring site from the impurity. Here, the bath particle moves to the left but the symmetric case is equally likely. The DW and ADW excitations can propagate in opposite directions (configuration 1) or the DW can propagate and the ADW form a stationary bound state with the impurity (configuration 2). From 2), the state can evolve for instance so that the DW and ADW propagate in the same direction (2 a) or the DW continues to the left and the ADW remains stationary, with the impurity oscillating between the two sites (2 b). In configurations 2 a) and 2 b), the DW excitation has moved to the left beyond the region drawn here.}
\label{fig:schematic}
\end{center}	
\end{figure*}

Instead of the simple description above, we find evidence that the DW-ADW pair is created right next to the impurity, in which case the motion of the excitations is restricted by energy conservation. Two distinct processes which can occur, and configurations to which the system can branch later, are illustrated in Fig. \ref{fig:schematic}. In one process, the DW and the ADW simultaneously propagate in opposite directions while the impurity stays at $j_0$ (configuration 1). In the other process, the impurity hops to the neighboring site (configuration 2). From configuration 2, the ADW can start propagating in the same direction as the DW, in which case the impurity will be trapped at $j_0 \pm 1$ (2a). Alternatively, the impurity and the ADW can remain in place and form a bound state while only the DW propagates (2b). If the impurity stays at $j_0$, the ADW cannot move to the left since this would require an additional energy $U$. This situation is different from creating the DW-ADW excitation far from the impurity as discussed above. Moreover, it is now possible for the impurity to hop within the effective double well formed by the ADW, associated with the kinetic energy~$J$. 

We find that instead of a kinetic energy contribution $\pm 4J$, the simulations are compatible with $\pm 3J$, as seen in Table~\ref{table:Uvalues}. We have located the limits of the resonance region, $U_{\min}$ and $U_{\max}$, by simulating the dynamics for different values of $V \gg J$, and different values of $U$ around $V$. The studied values of $U$ between which a clear decay in $\langle N_{\uparrow\downarrow}(t) \rangle$ starts and ends are indicated as intervals in Table~\ref{table:Uvalues}. These intervals are compared to $U_{\min}$ and $U_{\max}$ calculated by replacing the bounds of the bath excitation continuum $V \pm 4J$ by $V \pm 3J$. The agreement with $V \pm 3J$ may be related to the kinetic energy contribution from the impurity in addition to the bath excitations.

\begin{table}[h!]
\begin{center}
\caption{Values of $U$ and $V$ for which a decay of $\langle N_{\uparrow\downarrow} \rangle$ is reasoned in the discussion above, and found in the TEBD simulations.}
 \begin{tabular}{l c c c}
									&$V = 100 J$		&$V = 20 J$		&$V = 10 J$ \\
 \hline
$U_{\min} = \sqrt{(V - 3 J)^2 - 16 J^2}$	&96.9				&16.5			&5.7	\\
TEBD: $U_{\min} \in$				&$[96.8, 97.5]$		&$[17, 18]$		&$[6, 8]$	\\
$U_{\max} = V + 3 J$	&103				&23				&13		\\
TEBD: $U_{\max} \in$				&$[102.5, 103.2]$	&$[22, 23]$		&$[12, 13]$	\\
\end{tabular}
\label{table:Uvalues}	
\end{center}
\end{table}

\section{The time evolution of the bath}
\label{section:density_correlation}

The striped pattern of $\langle n_{j\uparrow}(t) \rangle - \langle n_{j\uparrow}(0) \rangle$ in Figs. \ref{fig:densities} and \ref{fig:densities_supplemental} corresponds to inverting the positions of empty and occupied sites in the bath, which is consistent with a DW or an ADW excitation propagating away from the center. At this time scale, the impurity on the other hand stays confined to the center of the lattice. The right top panel of Fig. \ref{fig:densities} shows that for $U = 90J$, the impurity does not move from the central site. This is consistent with the very small change in the density of bath fermions in the left top panel, which indicates that excitations are essentially not created. For $U = 100J$, the density changes in the bath are larger by an order of magnitude and the impurity has a considerable probability ($\langle n_{j_0 \pm 1 \downarrow} \rangle \approx 0.2$) to move to the neighboring site. While for a DW and an ADW propagating in opposite directions, the impurity stays at the central site, in the other states presented in Fig. \ref{fig:schematic}, the neighboring site will be occupied by the impurity with some probability.

Since the expectation value of the density is an average over all states with excitations propagating in either direction, it does not give information on whether both the DW and ADW move simultaneously or if only one of them moves and the other one stays fixed. Instead, one can study the density correlation $\bra{\psi(t)} \Delta n_{i\uparrow} \Delta n_{-i\uparrow} \ket{\psi(t)}$, where $\Delta n_{i\uparrow} = n_{i\uparrow} - \bra{\psi(0)} n_{i\uparrow} \ket{\psi(0)}$ and $i$, $-i$ are indices from the center of the lattice. If the DW and ADW propagate in opposite directions, the density difference $\langle \Delta n_{i\uparrow} \rangle$ will be 1 or -1 symmetrically on either side of $j_0$, giving a correlation that is maximally one. If only one excitation propagates, e.g. in the positive $i$ direction, $\langle \Delta n_{-i\uparrow} \rangle$ will stay zero and $\langle \Delta n_{i\uparrow} \Delta n_{-i\uparrow} \rangle = 0$. The correlation will also be zero if the DW and ADW propagate in the same direction, or if there are no excitations. Detailed examples of the calculation of $\langle \Delta n_{i\uparrow} \Delta n_{-i\uparrow} \rangle$ are shown in Appendix \ref{app:density_correlation}. Figure \ref{fig:density_correlation} shows $\langle \Delta n_{i\uparrow} \Delta n_{-i\uparrow} \rangle$ at different time steps for $U = V = 100J$. The correlation at $i = 1$ seems to saturate to approximately 0.5. This implies that the system is in a superposition where in addition to the oppositely moving DW and ADW, there are other states present for which the correlator is zero. Note that the physical picture of DW and ADW excitations can also help to explain previous results on the interaction-independent velocity of density correlations in Mott insulators \cite{Muramatsu}.

\begin{figure}[h!]
\begin{center}
	\includegraphics[width=\linewidth]{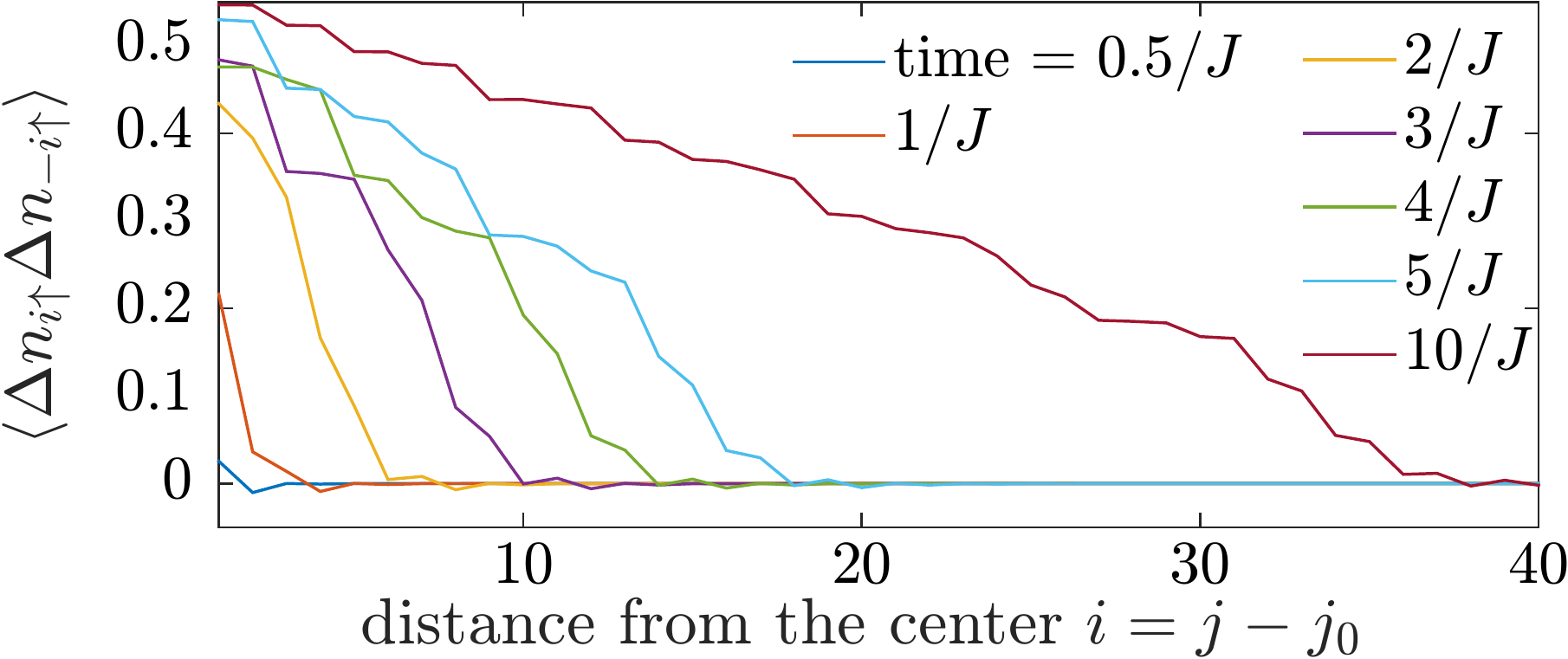}
	\caption{The density correlation $\bra{\psi(t)} \Delta n_{i\uparrow} \Delta n_{-i\uparrow} \ket{\psi(t)}$ for $U = V = 100J$ at different time steps shows that the bath evolves in a superposition of configurations (see Appendix \ref{app:density_correlation}).}
\label{fig:density_correlation}
\end{center}	
\end{figure}

\section{Experimental realization with ultracold dipolar gases}
\label{section:experiments}

Polar KRb molecules and Rydberg atoms in optical lattices have recently been used for realizing the spin-exchange interaction between nearest neighbors \cite{KRb_Jin2013, KRb_Jin2014}, and magnetic atoms have been employed for realizing a $t-J$-like model \cite{Laburthe-Tolra} and the extended Bose-Hubbard model \cite{EBHM_Ferlaino2015}. The magnitude of the nearest-neighbor interactions between the magnetic atoms ranges from zero to approximately 2 in units of the tunneling rates in these experiments \cite{Laburthe-Tolra, EBHM_Ferlaino2015}, but it is tunable by the lattice spacing and depth. The on-site interaction can be controlled by Feshbach resonances. The energy gap between the lowest and next-lowest energy bands is made an order of magnitude larger than the tunneling energies and interactions by tuning the lattice depth, preventing excitations to higher bands. In the $t-J$ experiment \cite{Laburthe-Tolra}, the largest band gap in units of the tunneling rate is in the $z$ direction, $\omega_z \approx 57 J_z$ and the largest nearest-neighbor interaction quoted is approximately $1.7 J_z$, where $J_z = 3$ Hz. A system with interactions close to $10 J$ and a band gap in this range could still be reasonably treated in the single-band approximation. This magnitude of interactions would be sufficiently high to observe the excitation dynamics studied here.

The tunneling and interaction energies are dependent on the lattice parameters; in particular, $\frac{V}{J}$ can be tuned by changing the lattice depth \cite{Cugliandolo}. In the extended Bose-Hubbard model experiment \cite{EBHM_Ferlaino2015}, lattice depths $(s_x, s_y, s_z) = (15, 15, 15)$ in units of the recoil energies correspond to $V \approx J = 27$~Hz in the $(x, y)$ plane. The effect of the nearest-neighbor interaction is already seen in the energy gap of the Mott insulator state. We calculate, using a Gaussian approximation for the Wannier functions, that a value $V \approx 10 J$ where $J \approx 2.7$~Hz could be reached with lattice depths around $(s_x, s_y, s_z) = (24, 24, 24)$ with the same laser wavelengths (see Appendix \ref{app:ehm_parameters}). A sufficient duration of the experiment for the dynamics studied here is also realizable, since a coherent Bose-Einstein condensate can be preserved for around 1 s \cite{EBHM_Ferlaino2015}, corresponding to $2.7 \frac{1}{J}$. For these low frequencies, temperature effects would play a role and should be taken into account. Note that post-selection techniques such as in Ref. \cite{Fukuhara} could help in that respect. 

Even larger dipole-dipole interactions can be attained by combining magnetic atoms into molecules \cite{Molecules_Ferlaino} and with heteronuclear molecules which possess large electric dipole moments \cite{Madison, Zwierlein}. The stability of polar molecules against chemical reactions is still an issue to be solved and has been approached by confining the molecules to deep lattices which suppresses their tunneling \cite{KRb_Jin2013}. Immobile polar molecules could be used for realizing the bath studied here, a Mott insulator at half filling with nearest-neighbor interactions, since it can be mapped to an XXZ model with spin exchange $S_i^+ S_{i + 1}^- + S_i^- S_{i + 1}^+$ and Ising $S_i^z S_{i + 1}^z$ terms. The impurity is not included in such a mapping, and one could study the possibility of whether an impurity particle added to the system would create the excitation dynamics predicted in this work.

\section{Conclusions}
\label{section:conclusions}

In summary, we have studied the dynamics of an impurity in a bath where the nearest-neighbor interaction leads to a periodic structure, in contrast to previous studies with homogeneous baths. While we consider parameters for which the bath is in the Mott insulator state, away from half filling a similar system can have a Luttinger parameter $K < \frac{1}{2}$. The dynamics of impurities in this regime are yet unexplored, and our results are an indication that also this parameter area may reveal novel types of dynamical phenomena. We find that structuring crucially affects the types of excitations in the system. An impurity which is initially localized at a site occupied by a bath particle can create an excitation where a DW and an ADW propagate in opposite directions. Since the ADW consists of two empty sites however, another excitation occurs where the impurity is coupled with an ADW and only a DW propagates. These new dynamical phenomena highlight that rich physics can emerge from a structured bath, and open interesting perspectives, for instance, to experiments on ultracold dipolar gases.

\begin{acknowledgments}
This work was supported by the Academy of Finland through its Centres of Excellence Programme (2012-2017) and under Project Nos. 263347, 251748, and 272490, and by the European Research  Council (ERC-2013-AdG-340748-CODE). A.-M. V. acknowledges financial support from the Vilho, Yrj\"o and Kalle V\"ais\"al\"a Foundation. Computing resources were provided by CSC--the Finnish IT Centre for Science and the Aalto Science-IT Project. This work was supported in part by the Swiss NSF under Division II and by the ARO-MURI Non-equilibrium Many-body Dynamics grant (W911NF-14-1-0003).
\end{acknowledgments}

\appendix
\section{Superlattice model for $U \ll V$ and $V \rightarrow \infty$}
\label{app:superlattice}

In the case $U \gg J$, $U \ll V$, and $V \rightarrow \infty$, the impurity problem is equivalent to a free particle in a superlattice with potential difference $U$ between alternating sites. When there is a higher potential on odd sites, one can write the number of doubly occupied sites as $\langle N_{\uparrow\downarrow}(t) \rangle
= \langle N_{\text{odd}}(t) \rangle = \sum_{j\text{ odd}} \bra{\psi(t)} c_j^{\dagger} c_j \ket{\psi(t)}$. The time evolution can be solved analytically, which allows to compare the results obtained for more realistic values of $U$ and $V$ to a perfectly rigid lattice. The Hamiltonian can be written as $H = H_J + H_U$, where
\begin{align*}
H_U = \frac{U}{2} \sum_j \left[ 1 - (-1)^{j} \right] c_{j}^{\dagger} c_j,
\end{align*}
and $H_J = -J \sum_{j\sigma} (c_{j\sigma}^{\dagger} c_{j+1\sigma} + \text{h.c.})$. Transforming $c_j = \frac{1}{\sqrt{L}} \sum_k e^{i k j} c_k$ gives
\begin{align*}
H_U &= \frac{U}{2} \sum_k c_k^{\dagger} c_k  - \frac{U}{2} \sum_j e^{i \pi j} \sum_{k, k'} e^{-i (k - k') j} c_k^{\dagger} c_{k'},
\end{align*}
where the first term is an energy offset and the second one can be written
\begin{align*}
- \frac{U}{2} \sum_j \sum_{k, k'} e^{-i (k - k' - \pi) j} c_k^{\dagger} c_{k'} 
&= - \frac{U}{2} \sum_{k, k'} \delta_{k, k' + \pi \text{(mod 2 $\pi$)}} c_k^{\dagger} c_{k'} \\ 
&= - \frac{U}{2} \sum_k c_{k + \pi \text{(mod 2 $\pi$)}}^{\dagger} c_{k}.
\end{align*}
We use periodic boundary conditions, which do not affect the result when the impurity is far from the edges of the lattice. A convenient way to diagonalize the Hamiltonian is to choose a new unit cell of length 2 (see \cite{Berthod}), which reduces the Brillouin zone from BZ ($k = -\pi + \frac{2 \pi n}{L}, n = 1, \cdots, L$) to BZ' ($k = -\frac{\pi}{2} + \frac{2 \pi n}{L}, n = 1, \cdots, \frac{L}{2}$). One can replace the operators $c_k$ with new operators $\alpha_k$ and $\beta_k$ defined in BZ',
\begin{align*}
\alpha_k &\coloneqq c_k, &k \in [-\frac{\pi}{2}, \frac{\pi}{2}], \\
\beta_{k - \pi} &\coloneqq c_k, &k \in [\frac{\pi}{2}, \pi], \\
\beta_{k + \pi} &\coloneqq c_k, &k \in [-\pi, -\frac{\pi}{2}].
\end{align*}
The Hamiltonian
\begin{align*}
H = \sum_{k \in [-\frac{\pi}{2}, \frac{\pi}{2}]} 
\begin{pmatrix}
\alpha_k^{\dagger}	&\beta_k^{\dagger}
\end{pmatrix}
\begin{pmatrix} 
\epsilon(k) + \Delta 	& -\Delta \\ 
- \Delta 				& -\epsilon(k) + \Delta
\end{pmatrix}
\begin{pmatrix}
\alpha_k	\\
\beta_k
\end{pmatrix},
\end{align*}
where $\epsilon(k) = -2J\cos(k)$ and $\Delta = \frac{U}{2}$, can be diagonalized by a Bogoliubov transformation
\begin{align*}
\gamma_{k -} = u_k \alpha_k - v_k \beta_k, \\
\gamma_{k +} = v_k \alpha_k + u_k \beta_k
\end{align*}
with the coefficients
\begin{align*}
u_k &= \frac{1}{\sqrt{2}} \sqrt{1 - \frac{\epsilon(k)}{\sqrt{\epsilon^2(k) + \Delta^2}}}, \\
v_k &= \frac{1}{\sqrt{2}} \sqrt{1 + \frac{\epsilon(k)}{\sqrt{\epsilon^2(k) + \Delta^2}}}.
\end{align*}
The diagonal elements are $\Delta + E_{k \pm}$, where
\begin{align}
E_{k \pm} = \pm \sqrt{\epsilon^2(k) + \Delta^2}.
\label{eq:energy_bands}
\end{align}

The wavefunction at time $t = 0$ can be written
\begin{align*}
\ket{\psi(0)} &= c_{j_0}^{\dagger} \ket{0} = \frac{1}{\sqrt{L}} \sum_{k \in BZ} e^{-i k j_0} c_k^{\dagger} \ket{0} \\
&= \frac{1}{\sqrt{L}} \sum_{k \in BZ'} e^{-i k j_0} (\alpha_k^{\dagger} + (-1)^{j_0} \beta_k^{\dagger}) \ket{0} \\
&= \frac{1}{\sqrt{L}} \sum_{k \in BZ'} e^{-i k j_0} [(u_k - (-1)^{j_0} v_k) \gamma_{k -}^{\dagger} + \\
& ((-1)^{j_0} u_k + v_k) \gamma_{k +}^{\dagger}]
\ket{0}, 
\end{align*}
and at time $t$,
\begin{align*}
\ket{\psi(t)} &= e^{-i H t} \ket{\psi(0)} \\
&= \frac{1}{\sqrt{L}} e^{-i \Delta t} \sum_{k \in BZ'} e^{-i k j_0} \times \\
&[(u_k - (-1)^{j_0} v_k) e^{-i E_{k -} t} \gamma_{k -}^{\dagger} + \\
&((-1)^{j_0} u_k + v_k) e^{-i E_{k +} t} \gamma_{k +}^{\dagger}] \ket{0} \\
&= \frac{1}{\sqrt{L}} e^{-i \Delta t} \sum_{k \in BZ'} e^{-i k j_0} (A_k^* \alpha_k^{\dagger} + B_k^* \beta_k^{\dagger}) \ket{0},
\end{align*}
where
\begin{align*}
A_k^* = [u_k - (-1)^{j_0} v_k] u_k e^{i E_k t} + \\
[(-1)^{j_0} u_k + v_k] v_k e^{-i E_k t}
\end{align*}
and
\begin{align*}
B_k^* = -[u_k - (-1)^{j_0} v_k] v_k e^{i E_k t} + \\
[(-1)^{j_0} u_k + v_k] u_k e^{-i E_k t}.
\end{align*}
In the exponents, $E_{k+}$ has been denoted by $E_k$. The number operator becomes
\begin{align*}
c_j^{\dagger} c_j &= \frac{1}{L} \sum_{q, q' \in BZ} e^{-i (q - q')j} c_q^{\dagger} c_{q'} \\
&= \frac{1}{L} \sum_{q, q' \in BZ'} e^{-i (q - q')j} [\alpha_q^{\dagger} \alpha_{q'} + \beta_q^{\dagger} \beta_{q'} +  \\
&(-1)^j \alpha_q^{\dagger} \beta_{q'} + (-1)^j \beta_q^{\dagger} \alpha_{q'}]
\end{align*}
and the expectation value
\begin{align}
\begin{split}
\bra{\psi(t)} c_j^{\dagger} c_j \ket{\psi(t)} &=
\frac{1}{L^2} \sum_{k, k' \in BZ'} e^{-i (k - k')(j - j_0)} [A_k A_{k'}^* + \\
&B_k B_{k'}^* + (-1)^j (A_k B_{k'}^* + B_k A_{k'}^*)] \\
&= \left| \frac{1}{L} \sum_{k \in BZ'} e^{i k (j - j_0)} f_k(j, j_0) \right|^2,
\end{split}
\label{eq:sum_squared}
\end{align}
where
\begin{align*}
f_k(j, j_0) &= A_k + (-1)^j B_k = \\
&\begin{cases*}
2 \cos(E_k t) - 4 i u_k v_k \sin(E_k t) & for $j_0, j$ odd \\
2 i (-u_k^2 + v_k^2) \sin(E_k t) & $j_0$ even, $j$ odd.
\end{cases*}
\end{align*}
Here, $j_0$ odd corresponds to an occupied site at $j_0$ and $j_0$ even to an empty site. The number of doubly occupied sites in the many-body model is now equal to the total occupation of the odd sites, $\langle N_{\uparrow \downarrow}(t) \rangle = \sum_{j\text{ odd}} \bra{\psi(t)} c_j^{\dagger} c_j \ket{\psi(t)}$. For $U \gg J$, $E_k \approx \frac{U}{2}$, and since the sum in eq. (\ref{eq:sum_squared}) contains terms of the type $\cos(E_k t)\cos(E_{k'}t) \approx \cos^2(\frac{U}{2})$, the oscillation frequency of $\langle N_{\uparrow \downarrow}(t) \rangle$ is close to $U$.

\section{Model for the resonance region}
\label{app:resonance_region}

\subsection{Energy of the impurity}

\begin{figure}[h!]
\begin{center}
	\includegraphics[width=0.7\linewidth]{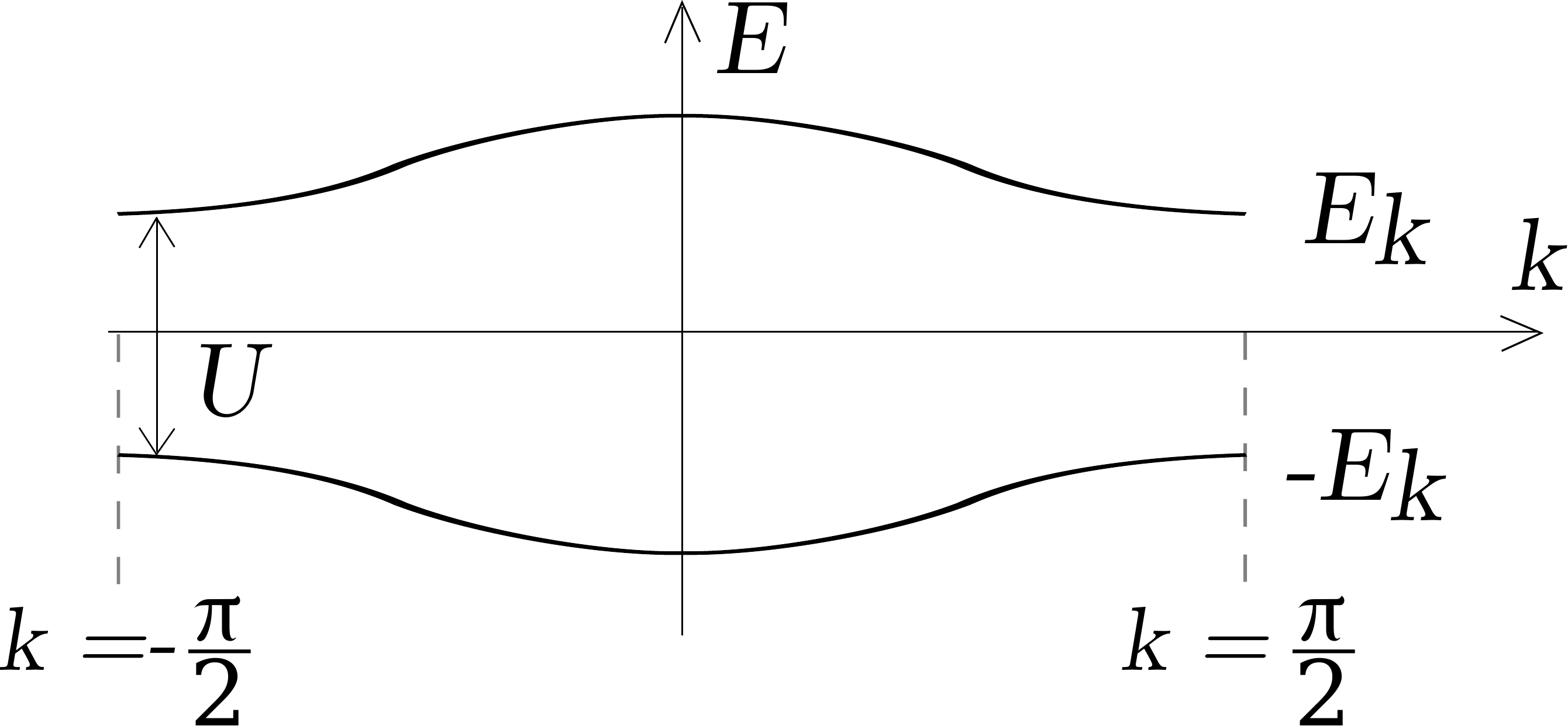}
	\caption{The energy bands of a free particle in a superlattice with potential difference $U$ between alternating sites.}
\label{fig:energy_bands}
\end{center}	
\end{figure}

A free particle in a superlattice with potential difference $U$ between alternating sites has the two energy bands $E_{k \pm} = \pm E_k$ given by eq. (\ref{eq:energy_bands}). They are illustrated in Fig. \ref{fig:energy_bands}. An impurity created at an occupied site is in the higher band, and moving to an empty site would correspond to a transition to the lower band. The change in the energy of the impurity would be 
\begin{align*}
\Delta E_{\text{imp.}} = E_{k^i} - (-E_{k^f}) &= \sqrt{4J^2\cos^2(k^i) + \left(\frac{U}{2} \right)^2} \\
&+ \sqrt{4J^2\cos^2(k^f) + \left(\frac{U}{2} \right)^2},
\end{align*}
which has the maximum $2 \sqrt{4J^2 + (\frac{U}{2})^2}$ and the minimum $U$. The initial and final momenta are denoted by $k^i$ and $k^f$. The energy released in such a transition would be absorbed by excitations created in the bath. 

\subsection{Energy of the bath}

The bath can be mapped to an XXZ Hamiltonian, which has a two-domain-wall excitation continuum $\omega \in [V - 4J, V + 4J]$ \cite{Shiba, Kolezhuk}. In our system, a bath fermion moving by one site corresponds to creating a domain wall (DW) with two neighboring sites occupied and an anti-domain wall (ADW) with two neighboring sites empty. Assuming the same excitation spectrum as for two domain walls \cite{Shiba, Kolezhuk}, this process has minimally the energy cost $V - 4J$ and maximally $V + 4J$. 

\subsection{Energy conservation}

A simple way to derive the minimum and maximum values of $U$ for which excitations can be created is to assume that the dispersion relation of the impurity is unchanged in the transition to the lower band. This means that the excitations in the bath are created far away and do not have an effect on the fixed superlattice potential close to the impurity. By energy conservation, $\Delta E_{\text{imp.}} = E_{DW+ADW}$. The lower limit for $U$ is now obtained when the change in the energy of the impurity has its maximum value and the DW-ADW excitation in the bath is created at the minimum energy $V - 4J$, 
\begin{align*}
2\sqrt{4J^2 + \frac{U_{\min}^2}{4}} = V - 4J.
\end{align*}
From this, one can solve $U_{\min} = \sqrt{(V - 4J)^2 - 16J^2}$. The upper limit is obtained when the energy change of the impurity has its minimum value $U$ and the excitation in the bath is created at the maximum energy $V + 4J$, 
\begin{align*}
U_{\max} = V + 4J.
\end{align*}

\section{Density correlation}
\label{app:density_correlation}

\begin{figure}[h!]
\begin{center}
	\includegraphics[width=\linewidth]{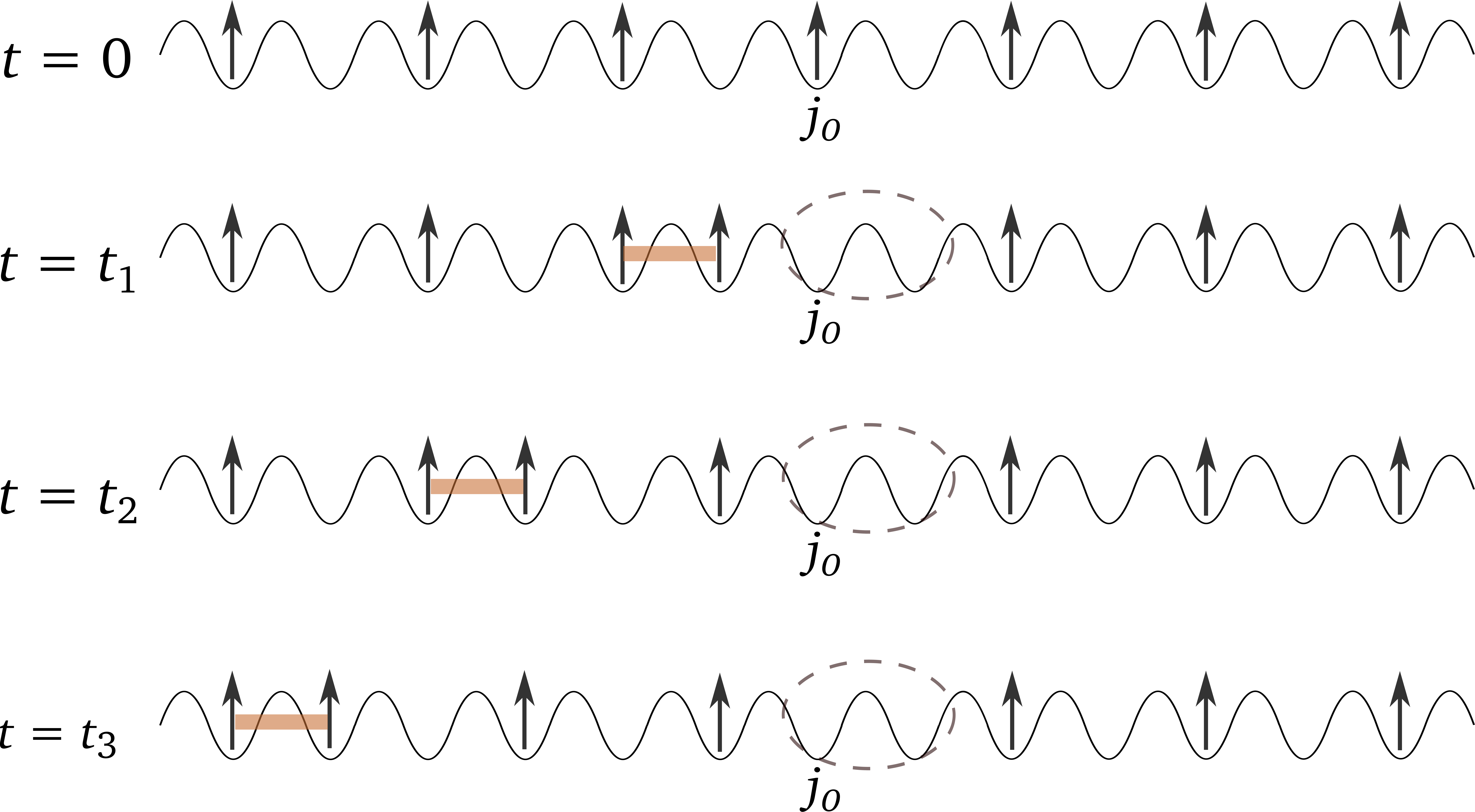}
	\caption{A DW propagating to the left and a stationary ADW.}
\label{fig:bound_ADW}
\end{center}	
\end{figure}

\begin{figure}[h!]
\begin{center}
	\includegraphics[width=\linewidth]{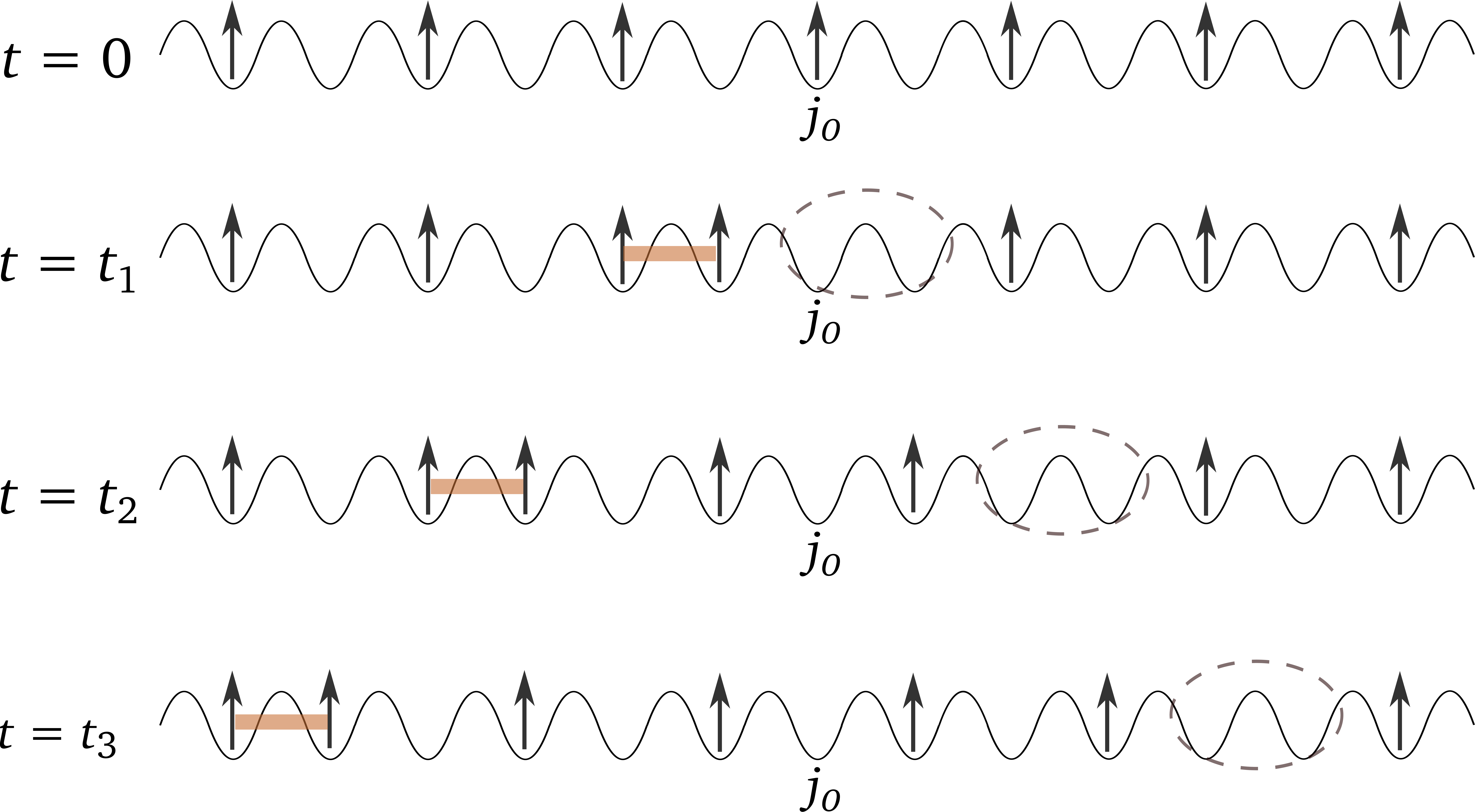} 
	\caption{A DW propagating to the left and an ADW propagating to the right.}
\label{fig:opposite_DW_ADW}
\end{center}	
\end{figure}

\begin{figure}[h!]
\begin{center}
	\includegraphics[width=0.5\linewidth]{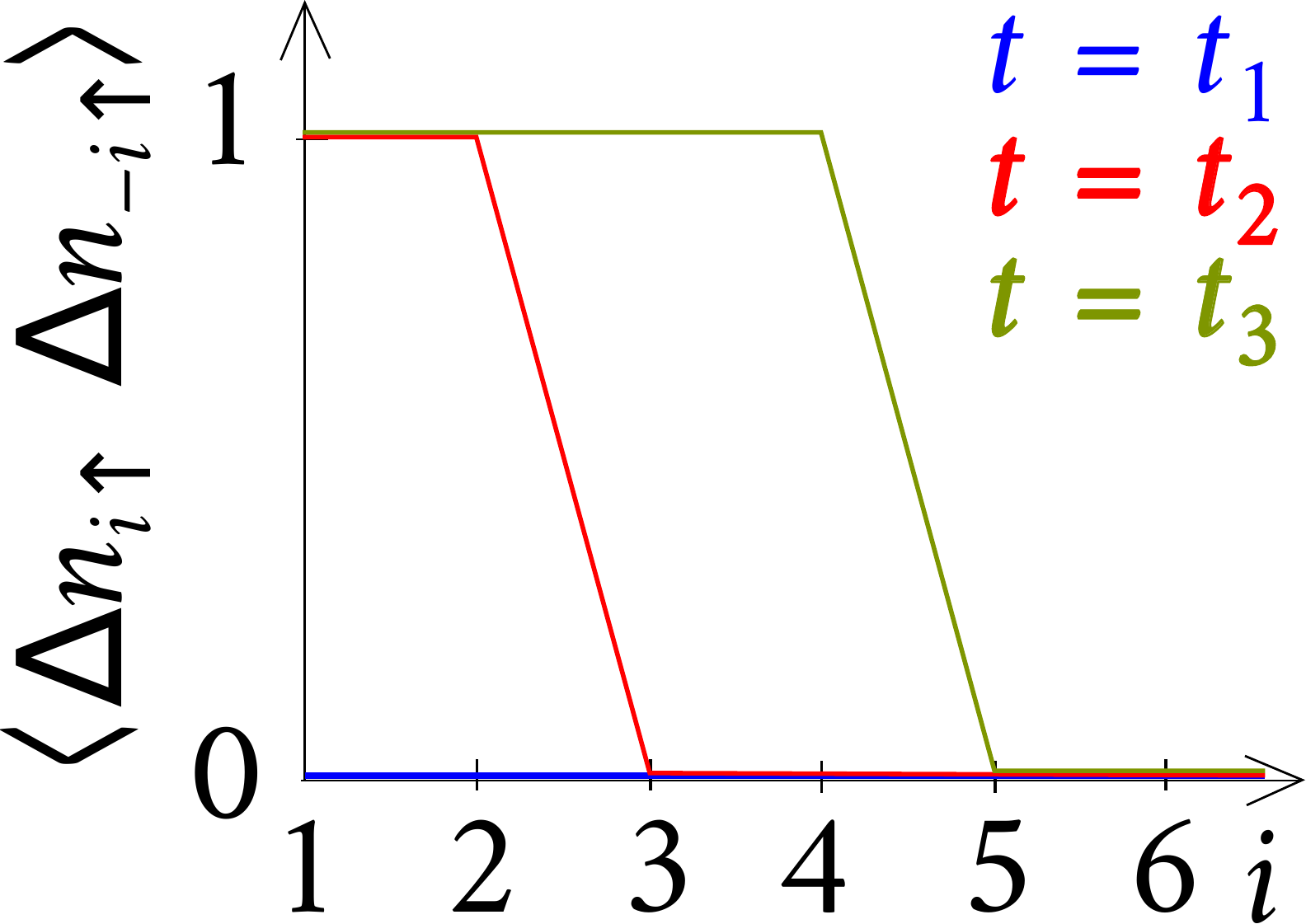}
	\caption{A schematic plot of the correlation $\bra{\psi(t)} \Delta n_{i\uparrow} \Delta n_{-i\uparrow} \ket{\psi(t)}$ at different time steps for the case of a DW propagating to the left and an ADW propagating to the right.}
\label{fig:correlation_schematic}
\end{center}	
\end{figure}

In this Section, we illustrate with two examples the meaning of the density correlation $\bra{\psi(t)} \Delta n_{i\uparrow} \Delta n_{-i\uparrow} \ket{\psi(t)}$, where $i$ is the distance to the center of the lattice $i = j - j_0$. Figures \ref{fig:bound_ADW} and \ref{fig:opposite_DW_ADW} depict two possible ways in which the bath configuration can evolve in time, similar to Fig. \ref{fig:schematic} of the main text. The impurity is not drawn for clarity. In Fig. \ref{fig:bound_ADW}, the DW excitation propagates to the left while the ADW excitation stays localized at the center of the lattice. In Fig. \ref{fig:opposite_DW_ADW}, the DW propagates to the left and the ADW to the right. In Tables \ref{table:bound_ADW} and \ref{table:opposite_DW_ADW}, the expectation value of the density change $\langle \Delta n_{j\uparrow}(t) \rangle = \bra{\psi(t)} n_{j\uparrow} \ket{\psi(t)} - \bra{\psi(0)} n_{j\uparrow} \ket{\psi(0)}$ is calculated for these two cases. When the state of the bath is described by a single configuration at any $t$ (and not a superposition of different configurations), as in these examples, $\bra{\psi(t)} \Delta n_{i\uparrow} \Delta n_{-i\uparrow} \ket{\psi(t)} = \langle \Delta n_{i\uparrow}(t) \rangle \langle \Delta n_{-i\uparrow}(t) \rangle$ and the value of the correlation can be read from Tables \ref{table:bound_ADW} and \ref{table:opposite_DW_ADW}. For example, Table \ref{table:opposite_DW_ADW} shows that $\bra{\psi(t_3)} \Delta n_{4\uparrow} \Delta n_{-4\uparrow} \ket{\psi(t_3)} = -1 \cdot (-1) = 1$. Figure \ref{fig:correlation_schematic} shows a schematic diagram of the density correlation as a function of $i$ at the different times corresponding to the configurations in Fig. \ref{fig:opposite_DW_ADW}. For the stationary ADW excitation of Fig. \ref{fig:bound_ADW}, the correlation is zero at all times. Figure \ref{fig:density_correlation} is the main text therefore shows that the system evolves in a superposition of different configurations.
\begin{table}[h!]
\begin{center}
\caption{The expectation values $\bra{\psi(0)} n_{j\uparrow} \ket{\psi(0)}$ and $\langle \Delta n_{j\uparrow}(t) \rangle$ for the configurations in Fig. \ref{fig:bound_ADW}.}
 \begin{tabular}{c | C{3.5mm} C{3.5mm} C{3.5mm} C{3.5mm} C{3.5mm} C{3.5mm} C{1cm} C{3.5mm} C{3.5mm} C{3.5mm} C{3.5mm} C{3.5mm} C{3.5mm}}
$j$ 			&1	&2	&3	&4	&5	&6	&$j_0 = 7$	&8	&9	&10	&11	&12	&13 \\
\hline
$\langle n_{j\uparrow} (0) \rangle$
				&1	&0	&1	&0	&1	&0	&1			&0	&1	&0	&1	&0	&1 \\
\hline
$\langle \Delta n_{j\uparrow}(t_1) \rangle$
				&0	&0	&0	&0	&0	&1	&-1			&0	&0	&0	&0	&0	&0 \\
$\langle \Delta n_{j\uparrow}(t_2) \rangle$
				&0	&0	&0	&1	&-1	&1	&-1			&0	&0	&0	&0	&0	&0 \\
$\langle \Delta n_{j\uparrow}(t_3) \rangle$
				&0	&1	&-1	&1	&-1	&1	&-1			&0	&0	&0	&0	&0	&0 \\					
\end{tabular}
\label{table:bound_ADW}	
\end{center}
\end{table}
\begin{table}[h!]
\begin{center}
\caption{The expectation values $\langle \Delta n_{j\uparrow}(t) \rangle = \bra{\psi(t)} n_{j\uparrow} \ket{\psi(t)} - \bra{\psi(0)} n_{j\uparrow} \ket{\psi(0)}$ for the configurations in Fig. \ref{fig:opposite_DW_ADW}.}
 \begin{tabular}{c | C{3.5mm} C{3.5mm} C{3.5mm} C{3.5mm} C{3.5mm} C{3.5mm} C{1cm} C{3.5mm} C{3.5mm} C{3.5mm} C{3.5mm} C{3.5mm} C{3.5mm}}
$j$ 			&1	&2	&3	&4	&5	&6	&$j_0 = 7$	&8	&9	&10	&11	&12	&13 \\
\hline
$\langle \Delta n_{j\uparrow}(t_1) \rangle$
				&0	&0	&0	&0	&0	&1	&-1			&0	&0	&0	&0	&0	&0 \\
$\langle \Delta n_{j\uparrow}(t_2) \rangle$
				&0	&0	&0	&1	&-1	&1	&-1			&1	&-1	&0	&0	&0	&0 \\
$\langle \Delta n_{j\uparrow}(t_3) \rangle$
				&0	&1	&-1	&1	&-1	&1	&-1			&1	&-1	&1	&-1	&0	&0 \\					
\end{tabular}
\label{table:opposite_DW_ADW}	
\end{center}
\end{table}

\section{Extended Hubbard model parameters}
\label{app:ehm_parameters}

In the extended Hubbard model in three dimensions, the hopping energy $J$ between neighboring sites is given by
\begin{align}
J = -\int d^3r \phi_j^*(\mathbf{r}) \left[ -\frac{\hbar^2}{2m} \nabla^2 + V_{\text{latt}}(\mathbf{r}) \right] \phi_{j + 1}(\mathbf{r}),
\label{eq:hopping}
\end{align}
where $\phi_j(\textbf{r})$ are the Wannier functions of the lowest band and $V_{\text{latt}}(\mathbf{r})$ is the lattice potential \cite{Cugliandolo}. The on-site interaction energy with the s-wave and dipole-dipole contributions is
\begin{align*}
U &= g\int d^3r |\phi_j(\textbf{r})|^4 \\
&+ \iint d^3r d^3r' |\phi_j(\textbf{r}')|^2 |\phi_j(\textbf{r})|^2 U(\textbf{r} - \textbf{r}')
\end{align*}
and the nearest-neighbor interaction is
\begin{align}
V = \iint d^3r d^3r'|\phi_j(\textbf{r}')|^2 |\phi_{j + 1}(\textbf{r})|^2 U(\textbf{r} - \textbf{r}').
\label{eq:NNI}
\end{align}
The dipole-dipole interaction $U(\mathbf{r})$ is given by 
\begin{align*}
U(\mathbf{r}) = \frac{C_{\text{dd}}}{4 \pi} \frac{1 - 3\cos^2 \theta}{r^3}
\end{align*}
for dipoles aligned in the $z$ direction with an angle $\theta$ between the dipole orientation and the relative location of the dipoles. The coupling constant is $C_{\text{dd}} = \mu_0 \mu^2$. The magnetic moment $\mu$ of for example Erbium is $7 \mu_B$. We use the lattice spacings $d_{x, y} = 266$ nm and $d_z = 532$ nm \cite{EBHM_Ferlaino2015}, fit the coefficients in eqs. (\ref{eq:hopping}) and (\ref{eq:NNI}) to the values of $J$ and $V$ in the $(x, y)$ plane given in \cite{EBHM_Ferlaino2015}, and calculate $J$ and $V$ using a different lattice depth.

\bibliographystyle{unsrt}
\addcontentsline{toc}{section}{Bibliography}
\bibliography{bibfile,totphys_overcomplete}

\end{document}